\documentclass[fleqn,twoside]{article}
\usepackage{espcrc2}
\usepackage{epsfig}

\def\be{\begin{equation}}
\def\ee{\end{equation}}
\def\bea{\begin{eqnarray}}
\def\eea{\end{eqnarray}}

\title{Recent results from $e^+e^- \to$ hadrons}

\author{S. I. Eidelman\address[BINP]{Budker Institute of Nuclear
    Physics, 11 Lavrentyev Ave., Novosibirsk 630090, Russia}}

\begin{document}

\begin{abstract}
New results on the low energy $e^+e^-$
annihilation into hadrons from Novosibirsk and Beijing are described.
Implications of the new measurements for the evaluation
of the hadronic contribution to the muon anomalous magnetic 
moment  are discussed.
\vspace{1pc}
\end{abstract}
\maketitle

\section{Introduction}

$e^+e^-$ annihilation into hadrons is one of the most important 
suppliers of experimental information on the quark interactions. At 
high energy it serves as a test of perturbative QCD whereas at low energies
it provides insight into nonperturbative effects in QCD as well
as valuable input to various phenomenological models describing
strongly interacting particles.

It became conventional to use the dimensionless quantity $R(s)$
to characterize the total cross section of $e^{+}e^{-} \to$ hadrons:

\be
\label{eq:R}
R=\sigma (e^{+}e^{-}\rightarrow hadrons)/
\sigma (e^{+}e^{-}\rightarrow \mu ^{+}\mu ^{-}).
\ee 

$R(s)$ is widely used for various calculations. 
Particularly, knowledge of $R(s)$ with high accuracy is required for
the evaluation of $a_{\mu}^{had,LO}$, the leading order 
hadronic contribution to the anomalous magnetic
moment of the muon  $a_{\mu}=(g_{\mu}-2)/2$ (see \cite{kino} and 
references therein) or the value of the fine structure constant 
at the $Z$ boson mass~\cite{EJ}.

$a_{\mu}$ known today to better than 1.0 ppm is one of the best measured
quantities in physics. 
The recently reported measurement of $a_{\mu}$  by the E821 collaboration 
at BNL \cite{E821} and a possible deviation of its result from 
the predictions of Standard Model (SM) \cite{DH,DEHZ,HMNT} has
generated numerous speculations (for a review and discussion 
see~\cite{cm01}). \\ 
   
Within SM, the uncertainty of the theoretical value of the leading 
order $a_{\mu}$   is dominated by the uncertainty of the hadronic 
contribution that can be calculated via the dispersion integrals.

\be
\label{eq:dint}
a_{\mu}^{had,LO} = 
\left( \frac{\alpha m_{\mu}}{3\pi} \right)^2
\int_{4m_\pi^2}^{\infty} \frac{R(s)\hat{K}(s)}{s^2}ds,
\ee  

\noindent
where the QED kernel $\hat{K}(s)$ is a smooth function of energy varying
from 0.63 at $s=4m_{\pi}^2$ to 1 at $s \to \infty$.

The precision of the $a_{\mu}^{had,LO}$ calculation depends on the 
approach used and varies from 1.34 ppm based on $e^+e^-$ data only
\cite{EJ} to 0.53 ppm if in addition $\tau$-lepton decay data as well as 
perturbative QCD and QCD sum rules are extensively used \cite{DH}. 
As it is clear from Eq.~\ref{eq:dint}, the major contribution 
to its uncertainty 
comes from the systematic error of the $R(s)$ measurement at low energies
($s<$ 2 GeV$^2$), which, in turn, is dominated by the 
systematic error of the measured cross section $e^+e^- \to \pi^+\pi^-$
or pion form factor $F_{\pi}$ directly related to it.
 
Assuming conservation of the vector current (CVC) and 
isospin symmetry, the spectral function of the  
$\tau^- \to {\rm X}^- \nu_{\tau}$ decay, where X$^-$ is a vector hadronic
state with I=1 can be related to the corresponding isovector
state X$^0$ produced in  $e^+e^-$ annihilation
\cite{tsai}. (Here X can be $2\pi, 4\pi, \omega \pi, \ldots$).
The detailed measurement of 
the spectral functions was provided by ALEPH \cite{ALEPH}, 
OPAL \cite{OPAL} and CLEO-II \cite{CLEO}. Comparison of the hadronic 
cross sections measured at $e^+e^-$ colliders with the spectral
functions of the corresponding $\tau^-$ decays provides a test
of CVC. If CVC holds with high accuracy,
$\tau$-lepton decay data can be also used to substantially improve the 
accuracy of the calculations mentioned above \cite{ADH}. Recent 
indications that the accuracy of CVC relations is not as good as 
believed for many years necessitates a careful reanalysis of such
estimations~\cite{e20,DEHZ}. 
Thus, new high precision measurements of the
cross section of $e^+e^- \to hadrons$ and particularly of the  
pion form factor as well as precise determinations 
of the hadronic mass spectra in $\tau$ lepton decays become   
extremely important.

\section{New Results from $e^+e^-$ Colliders}

\subsection{Experiments at VEPP-2M}\label{subsec:2.1}

Since 1974 the $e^+e^-$ collider VEPP-2M has been successfully
running in the Budker Institute of Nuclear Physics in Novosibirsk
in the c.m.energy range from the threshold of hadron production to
1400 MeV~\cite{vepp2m}. Its maximum  luminosity reached 
$\sim3\cdot 10^{30}$ cm$^{-2}$s$^{-1}$ at the $\phi$ meson energy.
In the last series of experiments two detectors (CMD-2  and SND)
installed at VEPP-2M collected about 30 pb$^{-1}$ each.

CMD-2 described in detail elsewhere \cite {cmd2} is a general purpose 
detector. Inside a superconducting solenoid with a field of 1 T there are
a drift chamber and proportional Z-chamber, both also used for the
trigger, and an endcap BGO electromagnetic calorimeter. Outside
there is a barrel CsI electromagnetic calorimeter and muon streamer
tube chambers. The main goal of CMD-2 is to perform a high precision 
measurement of the exclusive cross sections of various hadronic
channels of $e^+e^-$ annihilation and detailed studies of the low lying
vector mesons - $\rho, \omega$ and $\phi$.
 
SND described in detail elsewhere
\cite {snd} is a nonmagnetic detector with drift chambers for
tracking and a three layer NaI electromagnetic calorimeter.
Outside it there are a muon streamer tube  chamber and
plastic scintillators. The main goal of SND is to study
$\rho$, $\omega$ and $\phi$ decays as well as main hadronic channels.

Both experiments possess some special features 
making high precision measurements feasible:
\begin{itemize}
\item
large data samples due to the high integrated luminosity 
and large acceptance
\item
multiple scans of the same energy ranges to avoid possible
systematic effects; the step was 10 MeV in the c.m.energy for 
the continuum region and 1-2 MeV near the $\omega$ and $\phi$ peaks
\item
the absolute calibration of the beam energy using the resonance
depolarization method \cite{depol} reduces to a negligible
level a systematic error caused by an uncertainty in the energy 
measurement which can be significant for cross sections 
with strong energy dependence 
\item
good space and energy resolution lead to small background
\item
redundancy - unstable particles are independently detected via different
decay modes ($\omega \to \pi^+\pi^-\pi^0, \pi^0\gamma$; $\eta \to 
2\gamma, \pi^+\pi^-\pi^0, 3\pi^0, \pi^+\pi^-\gamma$)
\item
detection efficiencies and calorimeter response are studied by using 
"pure" experimental data samples rather than Monte Carlo events, 
e.g. more than 20 million $\phi$ meson decays per detector can be used 
for that purpose. 
\end{itemize}

New results are available on most of the hadronic channels. 

The process $e^+e^- \to \pi^+\pi^-$ is particularly important for
various applications because of its large cross section at low
energies. This reaction has been extensively studied 
before~\cite{dm1,OLYACMD,dm2}. The most precise pion form factor 
data were obtained in late 70s -- early 80s by CMD and OLYA detectors
\cite{OLYACMD}. Their accuracy was limited by systematic
errors of the experiments, varying from 2\% to 15\% over the VEPP-2M
energy range. In the new  measurement with the CMD-2 detector more
than 2 million events of the process
$e^+e^- \to \pi^+\pi^-$ were detected  from 370 to 1380 MeV. 
Below 600 MeV separation of Bhabha and $\pi^+\pi^-$ 
events is performed by measuring their momentum. Above this energy
the energy deposition of the final particles in the calorimeter
has been used. The number of events of the reaction 
$e^+e^- \to \mu^+\mu^-$
was evaluated from QED which validity at these energies had been
verified before.

\begin{figure}
\begin{center}
\epsfig{figure=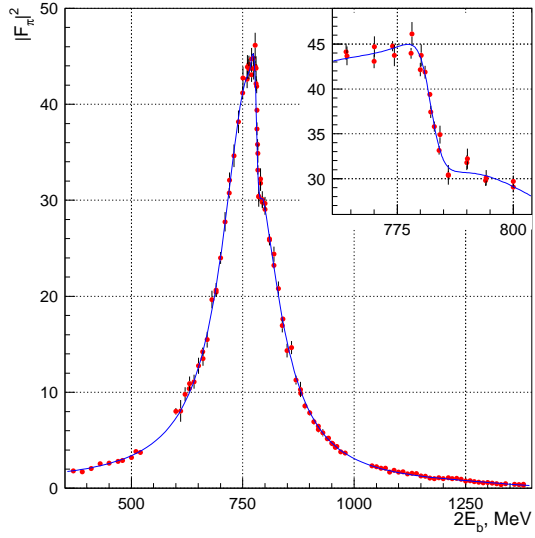,width=0.45\textwidth}
\caption{New data on the pion form factor.}
\label{fig:pipi}
\end{center}
\end{figure}

The systematic uncertainty of less than 0.6\% was achieved in the 
final analysis of the data set of about 114000 events collected in 
the energy range 610 to 960 MeV in 1994-1995 \cite{cmdpi}. 
Table~\ref{tab:syst} lists the dominant sources of the systematic error. 
Analysis is in progress for the rest of events and the expected 
systematic error ranges from 1\% to 3\% \cite{cmd2p}. 
Fig.~\ref{fig:pipi} shows  results of the pion form factor measurement 
coming from CMD-2.   

\begin{table}[t]
\caption{Main sources of systematic errors. \label{tab:syst} }
\vspace{0.4cm}
\begin{center}
\begin{tabular}{|l|c|}
\hline
Source & Contribution, \% \\
\hline 
Event separation& 0.2 \\
Radiative corrections & 0.4 \\
Detection efficiency & 0.2 \\
Fiducial volume &  0.2 \\
Correction for pion losses & 0.2 \\
Beam energy determination & 0.1 \\
\hline
Total & 0.6 \\
\hline 
\end{tabular}
\end{center}
\end{table}

There are still some weak points in this analysis: \\
\begin{itemize}
\item
it is important to determine directly the number of muons checking
thereby the correctness of the procedure of particle identification
as well as the normalization. To this end muon chambers can be used;  
\item
the knowledge of the radiative corrections now dominates
among possible sources of systematic uncertainties. It is 
necessary to check them using experimental events;
\item 
a correction for the radiation of a photon by final pions was
applied based on theoretical formulae that assumed pointlike
pions. One could try to test this assumption using experimental events 
with the radiation of a hard photon.
\end{itemize}
  
CMD-2 measured with high accuracy the main parameters of the
$\omega$ and $\phi $ mesons using their decays to $\pi^+\pi^-\pi^0$ 
\cite{cmdom,cmd3pi}, they also studied the $\phi$ meson in its
$K_SK_L$ decay mode \cite{cmdphi}. SND performed a high
precision measurement of three main decay modes of the $\phi$ 
meson in one experiment \cite{sndphi}. These studies allow a significant
improvement in the accuracy of the leptonic widths of the
$\omega$ and $\phi$ mesons.

SND also studied production of three pions above the $\phi$ and showed
that the energy dependence of the cross section is consistent with the
existence of at least one additional isoscalar resonance \cite{snd3pi}.
These conclusions are confirmed by preliminary results from CMD-2.   
The exact position of the resonance peak may be significantly lower
than that of the $\omega(1420)$.

Both detectors observed production of four pions. CMD-2 showed that 
in the energy range above the $\phi$ 
the $a_1(1260)^{\pm}\pi^{\mp}$ intermediate mechanism dominates 
in the $\pi^+\pi^-\pi^+\pi^-$ channel whereas both  
$a_1(1260)^{\pm}\pi^{\mp}$ and $\omega\pi$ contribute to 
the  $\pi^+\pi^-\pi^0\pi^0$ final state \cite{cmd4p}. 
The contribution of other possible intermediate states is small.
The collected
data sample includes about 60000 events and the systematic 
uncertainty of the total cross sections is less than 15\%. Below 1 GeV
CMD-2 reliably selected about 200 events of  
the reaction $e^+e^- \to \pi^+\pi^-\pi^+\pi^-$ and measured
the cross sections as low as about 50 pb near the $\rho$ 
peak \cite{cmd41}.
The measurement of the SND detector for which the data sample 
above the $\phi$ was about 80000 events and the systematic 
uncertainty ranged from 8 to 
20\%, confirmed the CMD-2 results on the production 
mechanisms \cite{snd4p}. 

\begin{figure}
  \begin{center}
      \epsfig{figure=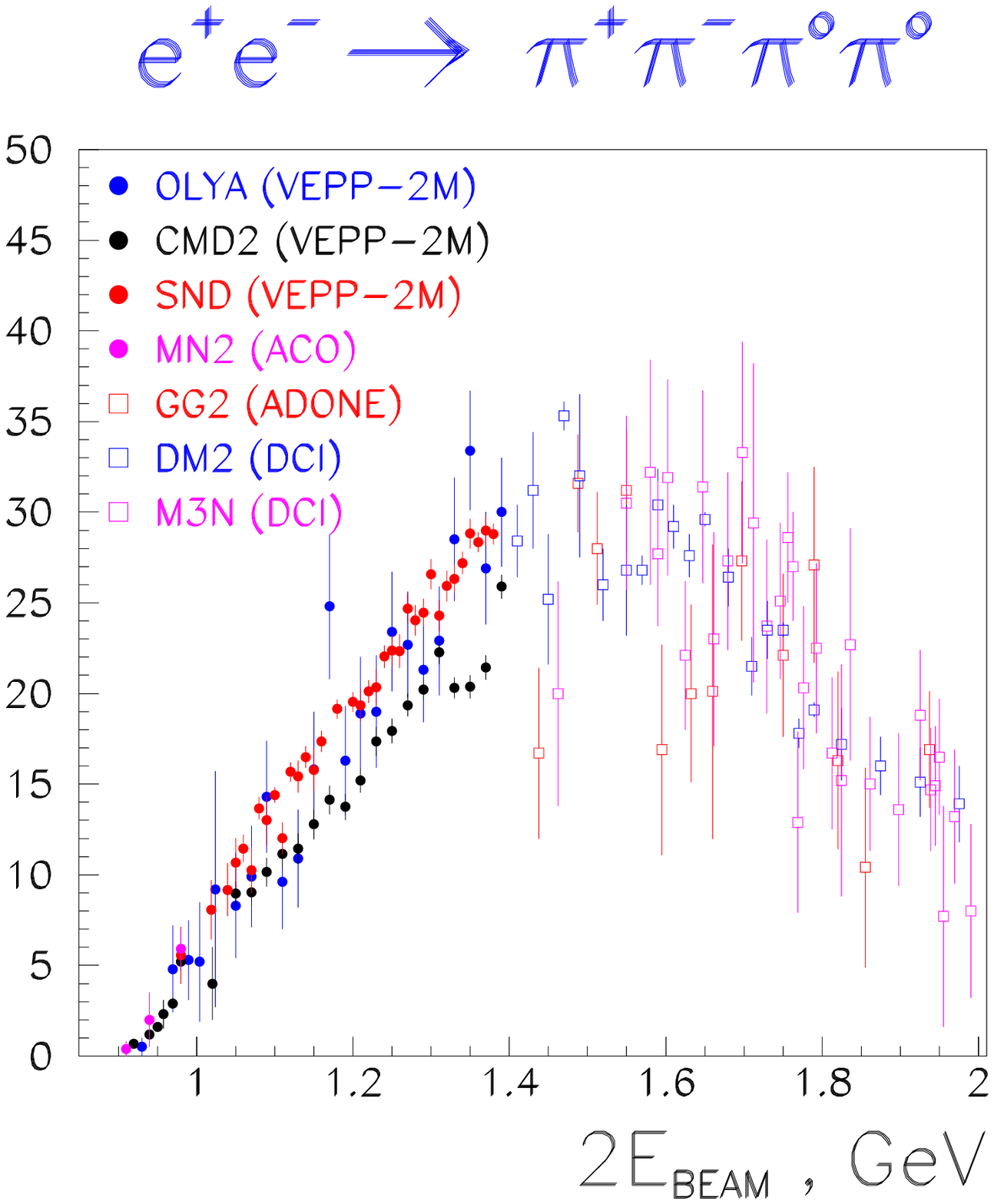,width=0.45\textwidth}
\caption{Cross section of the process $e^+e^- \to \pi^+\pi^-2\pi^0$.}
\label{fig:4pn}
\end{center}
\end{figure}

\begin{figure}
  \begin{center}
      \epsfig{figure=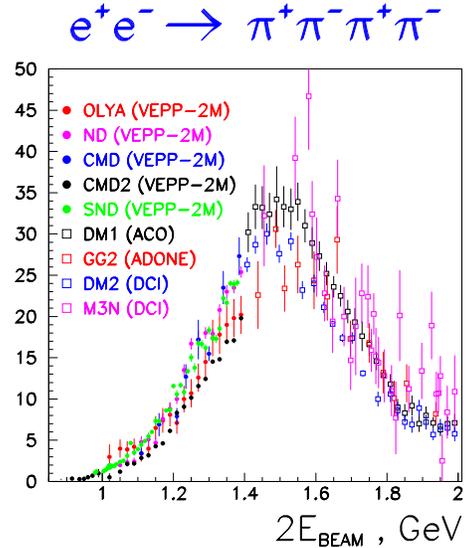,width=0.45\textwidth}
\caption{Cross section of the process $e^+e^- \to 2\pi^+2\pi^-$.}
\label{fig:4pc}
\end{center}
\end{figure}

However, in both 4$\pi$ channels the SND cross sections are higher than
those of CMD-2. The systematic uncertainties are still high and
their further analysis  is needed to clarify the picture. 
The corresponding cross sections are shown in 
Fig.~\ref{fig:4pn} and Fig.~\ref{fig:4pc} together with the results of 
the previous measurements at VEPP-2M, DCI and ADONE (for the references 
see \cite{cmd4p}).

Both detectors measured the cross section of the reaction 
$e^+e^- \to \omega\pi^0$ in the  $\omega \to \pi^0\gamma$ channel
collecting several thousand events each with the systematic error 
of 5\% for SND \cite{dru} and 8.5\% for CMD-2 \cite{krok}. Results
of both groups are consistent within systematic errors.

CMD-2 reliably observed multihadronic processes 
$e^+e^- \to \eta\pi^+\pi^-$
and $e^+e^- \to  \pi^+\pi^-\pi^+\pi^-\pi^0$
with the systematic accuracy of 15\% \cite{popov}.

CMD-2 has already published the results of the measurement
of the cross section for the process $e^+e^- \to K^0_SK^0_L$
~\cite{cmdkk} with (5--10)\% systematic uncertainty, which
are consistent with the preliminary measurements by SND~\cite{sndkk}.  
Analysis is in progress by CMD-2 for the $K^+K^-$  
final state produced at the $\phi$ meson and above it.

Both groups measured radiative decays of the $\omega$ and $\phi$
mesons \cite{sndn,cmdn}.

Thus, in the new experiments at the VEPP-2M collider in Novosibirsk
in the c.m.energy range from 0.37 to 1.38 GeV most of the hadronic 
reactions contributing to R have been measured  with much better accuracy
than before.
  

\subsection{R Measurement at BES}\label{subsec:2.2}

Until recently the energy range above 1.4 GeV was studied much worse
(see e.g. Figs.~\ref{fig:4pn},\ref{fig:4pc} where cross sections of the 
four pion production, the dominant process above 1 GeV, are shown below
the c.m.energy of 2 GeV). 
Despite numerous measurements of exclusive cross sections and R 
by various groups in Frascati, Orsay, DESY and SLAC the existing data
have big scatter and large systematic uncertainties ranging from
10\% to 25\%~\cite{DEHZ}. 

A real breakthrough occurred after recent experiments
with the BES detector at Beijing~\cite{bes} in which the total
cross section and R were thoroughly measured in the energy
range from 2 to 5 GeV. High statistics collected in this experiment
combined with the better acceptance than before and careful analysis of
the systematic uncertainties provided a basis for the significant
improvement of the accuracy of $R(s)$. Table~\ref{tab:r} illustrates 
the progress by
comparing some characteristics of the BES experiment with the
R measurement by the $\gamma\gamma2$ group at Frascati \cite{gg2}
in the energy range from 2.0 to 3.0 GeV.

\begin{table}[t]
\caption{Comparison of $\gamma\gamma2$ and BES measurements.\label{tab:r}}
\vspace{0.4cm}
\begin{center}
\begin{tabular}{|c|c|c|}
\hline
Detector &  $\gamma\gamma2$ & BES \\
\hline
$\sqrt{s}$, GeV  & 2.0 -- 3.1 & 2.0 -- 3.0 \\
Acceptance, \% & 19--23 & 50--68 \\
Syst.error, \% & 21 & 5.2--8.2 \\
$\int{Ldt},$nb$^{-1}$ & 130 & 990 \\
N$_{had}$ & 920 & 18500 \\
\hline
\end{tabular}
\end{center}
\end{table} 

\begin{figure}
\begin{center}
\epsfig{figure=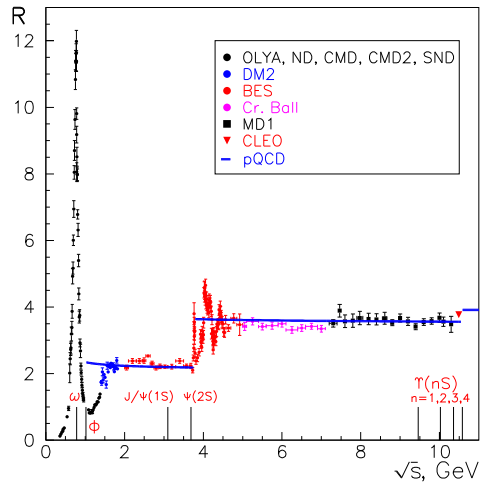,width=0.45\textwidth}
\caption{Total $e^+e^-$ cross section in terms of R}
\label{fig:R}
\end{center}
\end{figure}

In Fig.~\ref{fig:R} we show the experimentally measured value of R
as a function of energy below 10 GeV. Thresholds of $s\bar{s}$
and $c\bar{c}$ creation are obviously observed. The solid line showing 
the prediction of pQCD is in good agreement with the experimental 
observations. One can summarize that below 10 GeV (with the exception 
of the c.m.energy range from 1.4 to 2 GeV) the quantity R is known
with good accuracy.

\section{New $e^+e^-$ Data and $a_{\mu}^{had,LO}$}

Let us estimate the implications of the new  results on 
$e^+e^-$ cross sections discussed above 
for $a_{\mu}^{had,LO}$.  We'll start from  the 
contribution from the annihilation into two pions, which dominates
the hadronic contribution to $(g_{\mu}-2)/2$. 
To this end we compare its value in the energy range 610 to 960 MeV 
calculated from CMD-2 data only to that  based on the 
previous $e^+e^-$ measurements  \cite{dm1,OLYACMD}. Table~\ref{tab:g-2} 
presents results of the $a_{\mu}^{\pi\pi}$ calculations performed using
Eq.~(\ref{eq:dint}) and the 
direct integration of the experimental data. 
The method is straightforward and has been described
elsewhere \cite{EJ}. 
The first line of the Table~\ref{tab:g-2} (Old data) gives the result based 
on the data of OLYA, CMD and DM1 while the second one (New data) 
is obtained from the CMD-2 data only. 
The third line (Old + New) presents the 
weighted average of  these two estimates based on  
completely independent old and new datasets.
For convenience, we list separately statistical and systematic uncertainties
in the second column while the third one gives the total error 
obtained by adding them in quadrature. One can see that the 
estimate based on the CMD-2 data is in good agreement with that 
coming from the old data. It is worth noting that
the statistical error of the new measurement is slightly larger 
than  the systematic uncertainty. However, when the whole data sample
on two pion production collected by CMD-2 (more than one million 
events) is analyzed, one expects to significantly decrease the
statistical error above.
Because of the small systematic error of the new data, 
the uncertainty of the new result for $a_{\mu}^{\pi\pi}$ is 
almost three times better than the previous one. 
As a result, the combined value based on both old and new data is 
completely dominated by the CMD-2 measurement. 

One can now go further and estimate the full $a_{\mu}^{had,LO}$
by taking into account all previously available $e^+e^-$ experimental data
as well as the new data from Novosibirsk and Beijing described above
~\cite{DEHZ}. Similarly to Ref.~\cite{EJ}, only experimental data
will be used below 5 GeV whereas above that energy the predictions
of pQCD will be used. This approach is pretty safe since the energy
range above 5 GeV contributes only 1.5\% to  $a_{\mu}^{had,LO}$ and
the contribution to its uncertainty is negligible. The resulting value
is (684.7 $\pm$ 6.0 $\pm$ 3.6) $\cdot 10^{-10}$, where the first error 
is the total experimental error (including both statistical and systematic 
uncertainties) whereas the second one arises because of the
corrections for vacuum polarization and final state radiation.
The achieved accuracy of 7 $\cdot 10^{-10}$ is by a factor of more
than 2 better than the  previous one based on the $e^+e^-$ data 
only \cite{EJ}. This result is in good agreement with the recent
evaluation of  $a_{\mu}^{had,LO}$ in Ref.~\cite{HMNT} based on about
the same $e^+e^-$ data set, but using somewhat different method of
combining data of different experiments.   
   
\begin{table}[t]
\caption{Contributions of the $\pi\pi$ channel 
to (g-2)/2 \label{tab:g-2}  }
\vspace{0.4cm}
\begin{center}
\begin{tabular}{|l|c|c|}
\hline
Data  & a$_{\mu}^{\pi\pi}$, 10$^{-10}$ & Error, 10$^{-10}$ \\
\hline
Old   & 374.8 $\pm$ 4.1 $\pm$ 8.5 & 9.4 \\
New & 368.1 $\pm$ 2.6  $\pm$ 2.2 & 3.4 \\ 
\hline
Old+New & 368.9 $\pm$ 2.2 $\pm$ 2.3 & 3.2 \\ 
\hline
\end{tabular}
\end{center}
\end{table}    

Since E821 Collaboration plans to improve the accuracy, it is clear 
that further progress is also needed from the theoretical side.   
At the present time the analysis of the $\pi\pi$ data 
as well as other hadronic final states
in the whole energy range accessible to CMD-2 is in progress. 
Independent information is also  expected in close
future from other experiments studying
low energy $e^+e^-$ annihilation \cite{ret}. 
When all the above mentioned data are taken into account, one can expect
further improvement of the overall error  of $a_{\mu}^{had,LO}$
compared to the current one. 

It is very tempting to use an independent set of the $\tau$ lepton
data to improve the existing evaluations of the hadronic effects.
However, as we learned from the talk of
A.~H\"{o}cker at this Workshop~\cite{ah}, recent analysis revealed
the apparent incompatibility 
of $e^+e^-$ and  $\tau$ lepton spectral functions~\cite{DEHZ}. As a result,
the $\tau$ lepton based evaluation and the $e^+e^-$ based one 
are hardly consistent, so that their averaging doesn't improve the 
accuracy. More work is needed to clarify the reasons of this effect.

\section{Conclusions}

Thus, new experiments in Novosibirsk and Beijing considerably improved 
the accuracy of $R(s)$ in the energy ranges below 1.38 GeV and 
between 2 and 5 GeV allowing significant
improvement of the uncertainty of $a_{\mu}^{had,LO}$.

Precise tests of the relation between the $e^+e^-$ cross sections and
$\tau$ branching ratios will require better understanding of the
isospin symmetry breaking effects and radiative corrections.
Together with the more detailed analysis of systematic effects in 
both  $e^+e^-$ and $\tau$ lepton sectors, it should allow to
perform new precise evaluations of the hadronic corrections.
 
Further significant progress will become possible after new experiments
planned at Beijing, Cornell and Novosibirsk. Also promising looks
a possibility to study low energy   $e^+e^-$ annihilation by
the radiative return from the $\Upsilon(4S)$ or $\phi$ mesons
\cite{ret}.
 
\section{Acknowledgments}
The author is grateful to A.~Seiden and his colleagues from the
University of California, Santa Cruz  for an opportunity to present 
this talk and for the excellent organization of the Workshop.
Special thanks are due to A.E.~Bondar, A.~Czarnecki, M.~Davier, 
G.V.~Fedotovich, A.~H\"{o}cker, F.~Jegerlehner, P.P.~Krokovny,
L.M.~Kurdadze, W.J.~Marciano, A.I.~Milstein, and A.I.~Vainshtein  
for numerous stimulating discussions.


\end{document}